\pdfoutput=1
\documentclass[12pt,a4paper]{article}

\usepackage{ifthen} 
\newboolean{pdflatex}
\setboolean{pdflatex}{true} 

\newboolean{articletitles}
\setboolean{articletitles}{true} 

\newboolean{uprightparticles}
\setboolean{uprightparticles}{false} 

\def\paperauthors{LHCb collaboration} 
\def\paperasciititle{Measurement of the Lc to D0 production cross-section ratio in peripheral PbPb collisions} 
\def\papertitle{Measurement of the \Lc to \Dz production ratio in peripheral \PbPb \\ collisions at \sqsnn = 5.02 TeV} 
\def\paperkeywords{{High Energy Physics}, {LHCb}} 
\def\papercopyright{\the\year\ CERN for the benefit of the LHCb collaboration} 
\def\paperlicence{CC BY 4.0 licence}
\def\paperlicenceurl{https://creativecommons.org/licenses/by/4.0/}

\usepackage[top=1in, bottom=1.25in, left=1in, right=1in]{geometry}

\usepackage{microtype}
\usepackage{lineno}  
\usepackage{xspace} 
\usepackage{caption} 

\usepackage{graphicx}  
\usepackage{color}
\usepackage{colortbl}
\graphicspath{{./figs/}} 

\usepackage{amsmath} 
\usepackage{amssymb}
\usepackage{amsfonts}
\usepackage{upgreek} 

\newcommand*\patchAmsMathEnvironmentForLineno[1]{%
\expandafter\let\csname old#1\expandafter\endcsname\csname #1\endcsname
\expandafter\let\csname oldend#1\expandafter\endcsname\csname
end#1\endcsname
 \renewenvironment{#1}%
   {\linenomath\csname old#1\endcsname}%
   {\csname oldend#1\endcsname\endlinenomath}%
}
\newcommand*\patchBothAmsMathEnvironmentsForLineno[1]{%
  \patchAmsMathEnvironmentForLineno{#1}%
  \patchAmsMathEnvironmentForLineno{#1*}%
}
\AtBeginDocument{%
\patchBothAmsMathEnvironmentsForLineno{equation}%
\patchBothAmsMathEnvironmentsForLineno{align}%
\patchBothAmsMathEnvironmentsForLineno{flalign}%
\patchBothAmsMathEnvironmentsForLineno{alignat}%
\patchBothAmsMathEnvironmentsForLineno{gather}%
\patchBothAmsMathEnvironmentsForLineno{multline}%
\patchBothAmsMathEnvironmentsForLineno{eqnarray}%
}

\usepackage{hyperxmp}

\usepackage[pdftex,
            pdfauthor={\paperauthors},
            pdftitle={\paperasciititle},
            pdfkeywords={\paperkeywords},
            pdfcopyright={Copyright (C) \papercopyright},
            pdflicenseurl={\paperlicenceurl}]{hyperref}

\usepackage[colorinlistoftodos,textsize=scriptsize]{todonotes}

\usepackage[bottom,flushmargin,hang,multiple]{footmisc}

\usepackage[all]{hypcap} 

\usepackage{xspace} 
\usepackage{upgreek}

\def\lhcb   {\mbox{LHCb}\xspace}

\def\cms    {\mbox{CMS}\xspace}
\def\alice  {\mbox{ALICE}\xspace}

\def\velo   {VELO\xspace}

\def\ecal   {ECAL\xspace}
\def\hcal   {HCAL\xspace}
\def\MagUp {\mbox{\em Mag\kern -0.05em Up}\xspace}

\ifthenelse{\boolean{uprightparticles}}%
{

 \def\Ppi         {\ensuremath{\uppi}\xspace}

 \def\PDelta      {\ensuremath{\Delta}\xspace}                 
 \def\PXi         {\ensuremath{\Xi}\xspace}                 
 \def\PLambda     {\ensuremath{\Lambda}\xspace}                 
 \def\PSigma      {\ensuremath{\Sigma}\xspace}                 
 \def\POmega      {\ensuremath{\Omega}\xspace}                 
 \def\PUpsilon    {\ensuremath{\Upsilon}\xspace}
 \let\oldPi\Pi
 \def\PPi         {\ensuremath{\oldPi}\xspace}

 \def\PB      {\ensuremath{\mathrm{B}}\xspace}                 
                  
 \def\PD      {\ensuremath{\mathrm{D}}\xspace}

 \def\PK      {\ensuremath{\mathrm{K}}\xspace}

 \def\Pb      {\ensuremath{\mathrm{b}}\xspace}                 
 \def\Pc      {\ensuremath{\mathrm{c}}\xspace}

 \def\Pi      {\ensuremath{\mathrm{i}}\xspace}

 \def\Pp      {\ensuremath{\mathrm{p}}\xspace}

 \def\Ps      {\ensuremath{\mathrm{s}}\xspace}

 \def\thebaroffset{0.0em}
}
{

 \def\Ppi         {\ensuremath{\pi}\xspace}

 \mathchardef\PDelta="7101
 \mathchardef\PXi="7104
 \mathchardef\PLambda="7103
 \mathchardef\PSigma="7106
 \mathchardef\POmega="710A
 \mathchardef\PUpsilon="7107
 \mathchardef\PPi="7105
                  
 \def\PB      {\ensuremath{B}\xspace}                 
                  
 \def\PD      {\ensuremath{D}\xspace}

 \def\PK      {\ensuremath{K}\xspace}

 \def\Pb      {\ensuremath{b}\xspace}                 
 \def\Pc      {\ensuremath{c}\xspace}

 \def\Pi      {\ensuremath{i}\xspace}

 \def\Pp      {\ensuremath{p}\xspace}

 \def\Ps      {\ensuremath{s}\xspace}

 \def\thebaroffset{0.18em}
}
\newcommand{\offsetoverline}[2][\thebaroffset]{\kern #1\overline{\kern -#1 #2}}%

\makeatletter
\ifcase \@ptsize \relax
  \newcommand{\miniscule}{\@setfontsize\miniscule{4}{5}}
\or
  \newcommand{\miniscule}{\@setfontsize\miniscule{5}{6}}
\or
  \newcommand{\miniscule}{\@setfontsize\miniscule{5}{6}}
\fi
\makeatother

\DeclareRobustCommand{\optbar}[1]{\shortstack{{\miniscule (\rule[.5ex]{1.25em}{.18mm})}
  \\ [-.7ex] $#1$}}

\def\squark    {{\ensuremath{\Ps}}\xspace}

\def\cquark    {{\ensuremath{\Pc}}\xspace}

\def\bquark    {{\ensuremath{\Pb}}\xspace}

\def\pion   {{\ensuremath{\Ppi}}\xspace}

\def\pip    {{\ensuremath{\pion^+}}\xspace}
\def\pim    {{\ensuremath{\pion^-}}\xspace}

\def\kaon    {{\ensuremath{\PK}}\xspace}

\def\KorKbar {\kern \thebaroffset\optbar{\kern -\thebaroffset \PK}{}\xspace}

\def\Kp      {{\ensuremath{\kaon^+}}\xspace}
\def\Km      {{\ensuremath{\kaon^-}}\xspace}

\def\D       {{\ensuremath{\PD}}\xspace}

\def\DorDbar {\kern \thebaroffset\optbar{\kern -\thebaroffset \PD}\xspace}
\def\Dz      {{\ensuremath{\D^0}}\xspace}

\def\Dp      {{\ensuremath{\D^+}}\xspace}
\def\Dm      {{\ensuremath{\D^-}}\xspace}

\def\DpDm    {\ensuremath{\Dp {\kern -0.16em \Dm}}\xspace}

\def\B       {{\ensuremath{\PB}}\xspace}

\def\BorBbar {\kern \thebaroffset\optbar{\kern -\thebaroffset \PB}\xspace}

\def\Bd      {{\ensuremath{\B^0}}\xspace}

\def\BdorBdbar {\kern \thebaroffset\optbar{\kern -\thebaroffset \Bd}\xspace}

\def\Bs      {{\ensuremath{\B^0_\squark}}\xspace}

\def\BsorBsbar {\kern \thebaroffset\optbar{\kern -\thebaroffset \Bs}\xspace}

\def\Y#1S{\ensuremath{\PUpsilon{(#1S)}}\xspace}

\def\proton      {{\ensuremath{\Pp}}\xspace}

\def\Lz          {{\ensuremath{\PLambda}}\xspace}

\def\LorLbar     {\kern \thebaroffset\optbar{\kern -\thebaroffset \PLambda}\xspace}

\def\Lc          {{\ensuremath{\Lz^+_\cquark}}\xspace}

\def\BF         {{\ensuremath{\mathcal{B}}}\xspace}
\def\BR         {\BF}

\def\AT#1     {\ensuremath{A_{\mathrm{T}}^{#1}}\xspace}           

\def\C#1      {\ensuremath{\mathcal{C}_{#1}}\xspace}                       
\def\Cp#1     {\ensuremath{\mathcal{C}_{#1}^{'}}\xspace}                    
\def\Ceff#1   {\ensuremath{\mathcal{C}_{#1}^{\mathrm{(eff)}}}\xspace}        
\def\Cpeff#1  {\ensuremath{\mathcal{C}_{#1}^{'\mathrm{(eff)}}}\xspace}       
\def\Ope#1    {\ensuremath{\mathcal{O}_{#1}}\xspace}                       
\def\Opep#1   {\ensuremath{\mathcal{O}_{#1}^{'}}\xspace}                    

\newcommand{\nospaceunit}[1]{\ensuremath{\text{#1}}}       
\newcommand{\aunit}[1]{\ensuremath{\text{\,#1}}}       

\newcommand{\tev}{\aunit{Te\kern -0.1em V}\xspace}
\newcommand{\gev}{\aunit{Ge\kern -0.1em V}\xspace}
\newcommand{\mev}{\aunit{Me\kern -0.1em V}\xspace}
\newcommand{\kev}{\aunit{ke\kern -0.1em V}\xspace}
\newcommand{\ev}{\aunit{e\kern -0.1em V}\xspace}
 
\newcommand{\mevc}{\ensuremath{\aunit{Me\kern -0.1em V\!/}c}\xspace}
\newcommand{\gevc}{\ensuremath{\aunit{Ge\kern -0.1em V\!/}c}\xspace}
\newcommand{\mevcc}{\ensuremath{\aunit{Me\kern -0.1em V\!/}c^2}\xspace}
\newcommand{\gevcc}{\ensuremath{\aunit{Ge\kern -0.1em V\!/}c^2}\xspace}

\def\mum  {\ensuremath{\,\upmu\nospaceunit{m}}\xspace}

\def\mub{\ensuremath{\,\upmu\nospaceunit{b}}\xspace}

\newcommand{\chisq}{\ensuremath{\chi^2}\xspace}

\newcommand{\chisqip}{\ensuremath{\chi^2_{\text{IP}}}\xspace}

\def\gsim{{~\raise.15em\hbox{$>$}\kern-.85em
          \lower.35em\hbox{$\sim$}~}\xspace}
\def\lsim{{~\raise.15em\hbox{$<$}\kern-.85em
          \lower.35em\hbox{$\sim$}~}\xspace}

\def\sPlot{\mbox{\em sPlot}\xspace}

\def\sqs   {\ensuremath{\protect\sqrt{s}}\xspace}
\def\sqsnn {\ensuremath{\protect\sqrt{s_{\scriptscriptstyle\text{NN}}}}\xspace}
\def\pt         {\ensuremath{p_{\mathrm{T}}}\xspace}

\def\ptot       {\ensuremath{p}\xspace}

\def\tell1  {TELL1\xspace}
\def\ukl1   {UKL1\xspace}

\def\qgp    {\ensuremath{\mathrm{QGP}}\xspace}
\def\hf    {\ensuremath{\mathrm{HF}}\xspace}

\def\pPb {\ensuremath{\proton\mathrm{Pb}}\xspace}
\def\Pbp {\ensuremath{\mathrm{Pb}\proton}\xspace}
\def\PbPb {\ensuremath{\mathrm{Pb}\mathrm{Pb}}\xspace}

\newcommand{\npart}{\ensuremath{\langle \mathrm{N_{part}} \rangle}\xspace}

\def\effTot{\ensuremath{\epsilon_{\mathrm{tot}}}\xspace}
\def\effAcc{\ensuremath{\epsilon_{\mathrm{acc}}}\xspace}

\def\effTrk{\ensuremath{\epsilon_{\mathrm{rec\&sel}}}\xspace}
\def\effPID{\ensuremath{\epsilon_{\mathrm{PID}}}\xspace}

\newcommand{\eg}{\mbox{\itshape e.g.}\xspace}

\newcommand{\lhcborcid}[1]{\href{https://orcid.org/#1}{\hspace*{0.1em}\raisebox{-0.45ex}{\includegraphics[width=1em]{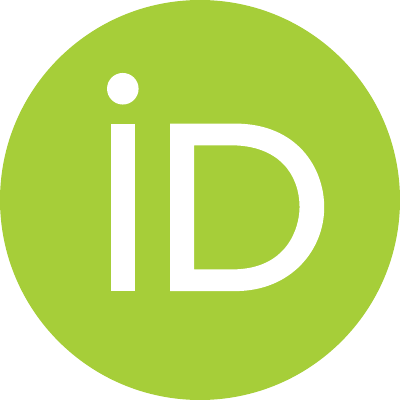}}}}

\usepackage{cite} 
\usepackage{mciteplus}

\usepackage{longtable} 

\usepackage{adjustbox} 

\begin{document}

\renewcommand{\thefootnote}{\fnsymbol{footnote}}
\setcounter{footnote}{1}


\begin{titlepage}
\pagenumbering{roman}

\vspace*{-1.5cm}
\centerline{\large EUROPEAN ORGANIZATION FOR NUCLEAR RESEARCH (CERN)}
\vspace*{1.5cm}
\noindent
\begin{tabular*}{\linewidth}{lc@{\extracolsep{\fill}}r@{\extracolsep{0pt}}}
\ifthenelse{\boolean{pdflatex}}
{\vspace*{-1.5cm}\mbox{\!\!\!\includegraphics[width=.14\textwidth]{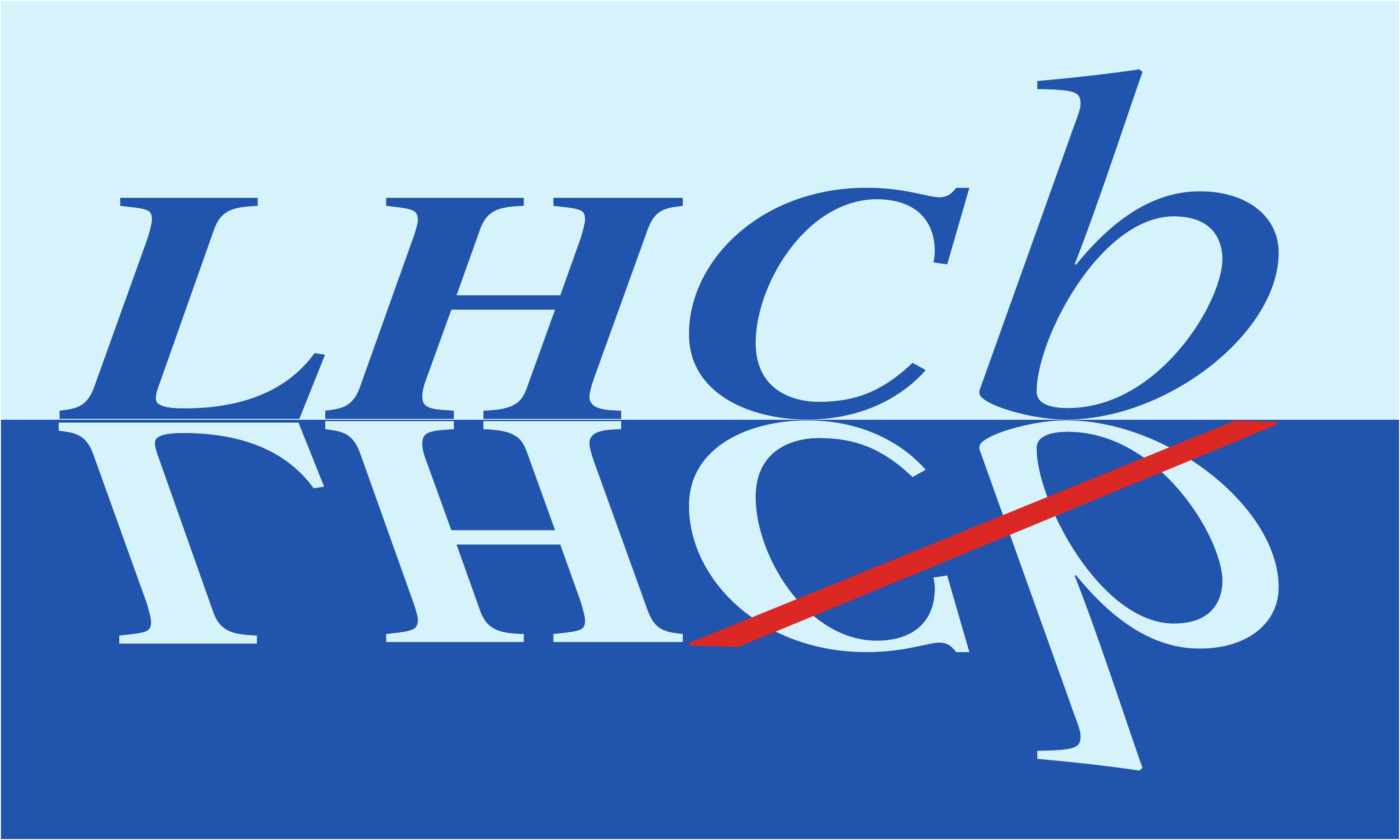}} & &}%
{\vspace*{-1.2cm}\mbox{\!\!\!\includegraphics[width=.12\textwidth]{figs/lhcb-logo.eps}} & &}%
\\
 & & CERN-EP-2022-148 \\  
 & & LHCb-PAPER-2021-046 \\  
 & & 13 October 2022 \\ 
 & & \\
\end{tabular*}

\vspace*{3.0cm}

{\normalfont\bfseries\boldmath\huge
\begin{center}
  \papertitle 
\end{center}
}

\vspace*{1.0cm}

\begin{center}
\paperauthors
\footnote{Authors are listed at the end of this paper.}
\end{center}

\vspace*{2.0cm}

\vspace{\fill}

\begin{abstract}
  \noindent
We report on a measurement of the \Lc to \Dz production ratio in PbPb collisions at \sqsnn = 5.02 TeV with the \lhcb detector in the forward rapidity region \mbox{ $2 <y< 4.5$}. The \Lc (\Dz) hadrons are reconstructed via the decay channel $\Lc\rightarrow p\Kp\pim$ \mbox{($\Dz \rightarrow \Km\pip$)} for $2<\pt<8$ \gevc and in the centrality range of about \mbox{65--90\%}. The results show no significant dependence on \pt, $y$ or the mean number of participating nucleons. They are also consistent with similar measurements obtained by  the \lhcb collaboration in pPb and Pbp collisions at \sqsnn = 5.02 TeV. The data agree well with predictions from PYTHIA in pp collisions at \sqs = 5 \tev but are in tension with predictions of the Statistical Hadronization model.
  
\end{abstract}


\begin{center}
  Published in JHEP06 (2023) 132
\end{center}

\vspace{\fill}

{\footnotesize 
\centerline{\copyright~\papercopyright. \href{\paperlicenceurl}{\paperlicence}.}}
\vspace*{2mm}

\end{titlepage}


\newpage
\setcounter{page}{2}
\mbox{~}
%
%
%
%


\renewcommand{\thefootnote}{\arabic{footnote}}
\setcounter{footnote}{0}

\cleardoublepage


\pagestyle{plain} 
\setcounter{page}{1}
\pagenumbering{arabic}


\section{Introduction}
\label{sec:Introduction}

Historically, heavy flavor (\hf) hadrons (e.g containing at least one  $c$ or $b$ quark) have been extensively used to study the deconfined state of hadronic matter, the Quark-Gluon Plasma (\qgp)~\cite{Braun_Munzinger_2016,PhysRevD.90.094503,Gossiaux:2019mjc}, in particular at the LHC and RHIC~\cite{Andronic_2016,Prino_2016,annurev-nucl-101918-023806}. Because their high masses exceed the QCD energy scale, charm and beauty quarks are produced at an early stage of the collision and experience the entire evolution of the colliding medium. Medium-induced energy loss has been studied by measuring the so-called nuclear modification factor ($R_{\rm AA}$), defined as the ratio of the production yield in nucleus-nucleus (AA) collisions to the one in \proton\proton collisions scaled by the number of binary nucleon-nucleon collisions. The measurements of the production cross-section, together with studies of the elliptic flow, indicate a strong interaction between \hf and the deconfined medium.
In addition, \hf hadrons provide a good laboratory to study hadronization. In particular, baryon-to-meson production ratios are of great interest as they are only sensitive to hadronization. Many measurements have been carried on in $e^+e^-$~\cite{CLEO:2004enr,Belle:2005mtx,BaBar:2002ncl} and \proton\proton collisions~\cite{LHCb-PAPER-2012-041,LHCb-PAPER-2015-041,LHCb-PAPER-2016-042} to measure fragmentation functions of heavy quarks into heavy hadrons, the latter being extensively used to describe hadron production at high transverse momentum (\pt). However, recent measurements of the \Lc to \Dz production ratio\footnote{If not stated otherwise, charge conjugation is assumed throughout the article.}  ($R_{\Lc/\Dz}$) challenge understanding of hadronization. At the LHC, \alice has measured the charmed baryon-to-meson ratio at mid-rapidity in \proton\proton, \pPb and \PbPb collisions at \sqs = 7\tev and $\mbox{\sqsnn = 5.02\tev}$, respectively~\cite{Acharya_2018,Acharya_2019,ALICE:2020wla,ALICE:2021bib}. An enhancement compared to predictions of the \proton\proton event generator PYTHIA 8~\cite{Christiansen2015} for the so-called MONASH tuning for \proton\proton data. This tuning does include a color reconnection (CR) mechanism but only at leading order approximation, and data can only be explained by the next-to-leading order approximation. In addition, another measurement of $R_{\Lc/\Dz}$ was carried out by ALICE versus the number of charged particles produced (i.e multiplicity) in \proton\proton collisions at \sqs = 13 \tev\cite{2022137065} at mid-rapidity. The results show an evolution of the \pt dependence of the ratio as the multiplicity increases, with a shift of the mean \pt distribution towards a higher value, while the \pt-integrated value of $R_{\Lc/\Dz}$ remains constant. The \PbPb result shows also a large enhancement of the ratio compared to \proton\proton and \pPb collisions, with a significant dependence of the \pt distribution with centrality. Similarly, \cms~\cite{2020} has performed the same measurement at mid-rapidity in \proton\proton and \PbPb collisions at \mbox{$\sqsnn = 5.02\tev$}, for \pt $>$ 5 \gevc and \pt $>$ 10 \gevc, respectively. Compatible results are found between the two data samples for the common range in \pt (i.e \pt $>$ 10 \gevc). In this case, the data are well described by PYTHIA 8 only when the CR mechanism beyond leading approximation is used. At the RHIC collider, the STAR collaboration has also measured $R_{\Lc/\Dz}$ at mid-rapidity in Au-Au collisions at \sqsnn = 200 \gev~\cite{Adam:2019hpq}. An enhanced baryon-to-meson ratio is found at low \pt compared to scaled \proton\proton collisions. It is worth mentioning that both \alice and STAR measurements in AA collisions can be described by a coalescence hadronization mechanism~\cite{PhysRevC.79.044905,Plumari_2018,Beraudo2022,PhysRevC.101.024909,Zhao:2018jlw,annurev-nucl-101918-023806} in which quarks can (re)combine with close-by partons in the \qgp to form hadrons. Other predictions based on the Statistical Hadronization Model (SHM)~\cite{He_2019}, using an augmented set of excited charm baryons based on the relativistic quark model (RQM), have successfully described the charmed baryon-to-meson ratio measured at RHIC and LHC. In this model, only the fragmentation of the heavy charm quark is taken into account using fragmentation functions. Finally, a last class of model based on transport equations, using both fragmentation and coalescence effect, can successfully reproduce the ALICE and RHIC data in nucleus-nucleus collisions~\cite{PhysRevLett.124.042301}.

As mentioned, \alice, \cms, and STAR measure a $R_{\Lc/\Dz}$ in heavy-ion collisions at mid-rapidity, with a raising trend from low to intermediate transverse momentum. Regardless of the theoretical model used to interpret the data, results at mid-rapidity point toward an effect that depends either on the charged particle multiplicity or the centrality. The latter can be interpreted as a dependence on the mean number of nucleons participating in the collision (\npart). On the other hand, LHCb has measured a ratio with no \pt dependence in \pPb collisions at \mbox {$\sqsnn = 5.02 \tev$} in the rapidity (\pt) range $2<y<4.5$ ($2<\pt<8\gevc$)~\cite{Aaij:2018iyy}. These results correspond to smaller $R_{\Lc/\Dz}$ values compared to measurements by other experiments and are fully compatible with predictions from cold nuclear matter effects~\cite{Eskola:2009uj,Kovarik:2015cma} within the HELAC-Onia approach~\cite{Lansberg_2016,arxiv.1507.03435,SHAO20132562}. The \lhcb results are yet to be compared with a coalescence-based model. These differences between mid and forward rapidity results indicate a strong rapidity dependence of this ratio and  motivate further studies to better improve the model predictions in different phase-space regions.

This paper presents the first measurement of $R_{\Lc/\Dz}$ production cross-section ratio in the forward rapidity region in peripheral \PbPb collisions at \sqsnn = 5.02\tev by the LHCb collaboration. The paper is organized as follows. Section~\ref{sec:Detector} briefly presents the detector and the data sample. Section~\ref{sec:analysis} describes the analysis steps, from the signal extraction to the estimation of efficiency corrections. The sources of systematic uncertainties are given in Sec.~\ref{sec:syst}. The results are presented and compared to theory predictions in Sec.~\ref{sec:results}, and conclusions are summarized in Sec.~\ref{sec:conclusion}.

\section{Detector and data selection}
\label{sec:Detector}

The \lhcb detector~\cite{LHCb-DP-2008-001,LHCb-DP-2014-002} is a single-arm forward
spectrometer covering the \mbox{pseudorapidity} range $2<\eta <5$,
designed for the study of particles containing \bquark or \cquark
quarks. The detector includes a high-precision tracking system
consisting of a silicon-strip vertex detector (\velo) surrounding the 
interaction region~\cite{LHCb-DP-2014-001}, a large-area silicon-strip detector located
upstream of a dipole magnet (UT) with a bending power of about
$4{\mathrm{\,Tm}}$, and three stations of silicon-strip detectors and straw
drift tubes~\cite{LHCb-DP-2017-001} placed downstream of the magnet.
The tracking system provides a measurement of the momentum, \ptot, of charged particles with
a relative uncertainty that varies from 0.5\% at low momentum to 1.0\% at 200\gevc.
The minimum distance of a track to a primary collision vertex (PV), the impact parameter (IP), 
is measured with a resolution of $(15+29/\pt)\mum$, with \pt in \gevc. Different types of charged hadrons are distinguished using information
from two ring-imaging Cherenkov detectors~\cite{LHCb-DP-2012-003}. 
Photons, electrons and hadrons are identified by a calorimeter system consisting of
scintillating-pad and preshower detectors (SPD), an electromagnetic (\ecal)
and a hadronic (\hcal) calorimeter. Muons are identified by a
system composed of alternating layers of iron and multiwire
proportional chambers~\cite{LHCb-DP-2012-002}.
The online event selection is performed by a trigger~\cite{LHCb-DP-2012-004}, 
which consists of a hardware stage, based on information from the calorimeter and muon
systems, followed by a software stage, which applies a full event
reconstruction.

The \PbPb collision data at \sqsnn = 5.02 \tev were recorded in 2018 and correspond to an integrated luminosity of about $210 \mub^{-1}$. Offline quality selections are applied on a run-by-run basis, based on the trend of trigger counts with time. At the hardware trigger stage, events containing \Lc or \Dz candidates are required to match the minimum bias (MB) trigger corresponding to a requirement of at least four SPD hits or a high-\pt muon (\pt $>$ 10 \gevc) or a minimal energy deposit in \hcal ($E_{\rm T} > 15$ \gevc). The events are required to have at least 15 VELO tracks in the backward direction and the number of  clusters (Nc) in the VELO should be at least 1000; these requirements  suppress contamination from both the Pb-gas\footnote{Simultaneously to \PbPb collisions, Neon gas was injected in the beam pipe near the interaction point, using the LHCb fixed-target SMOG system~\cite{LHCb-PAPER-2014-047}.} and ultra-peripheral collisions. The latter are defined as electromagnetic nucleus-nucleus interactions where the impact parameter ($b$) is greater than the sum of the nuclei radii. Finally, events are rejected if $N_c$ $\geq$ 10000, due to hardware limitations.

The \Lc (\Dz) candidates are reconstructed via the decay channel $\Lc\rightarrow p\Km\pip$ ($\Dz \rightarrow \Km\pip$), with a selection on transverse momentum  $2<\pt<8$ \gevc and rapidity $2<y<4.5$. Offline selections are applied to the candidates following the same strategy as in \pPb collisions~\cite{LHCb-PAPER-2018-021} to ensure a high signal significance and improve the purity of the \Dz and \Lc candidates. Pion, kaon, and proton tracks should match tracking and particle identification (PID) quality requirements. The \Lc (\Dz) decay products are required to have $\pt>400\mevc$ ($\pt > 500\mevc$) and $2 < y < 4.5$. The charm hadron lifetime is required to be less than 0.3 ps to reduce the fraction of non-prompt contribution coming from \bquark-hadron decay. The cosine of the direction angle between the candidate's momentum and the vector between the PV and the candidate's decay vertex, is required to be larger than $0.9998$. In addition, a fiducial cut around the beams' collision point is applied based on the PV of the \Lc and \Dz candidates.

Simulated \PbPb collisions at \sqsnn = 5.02\tev with full event reconstruction are used to evaluate efficiencies. The \Lc and \Dz candidates are generated with PYTHIA 8~\cite{Sj_strand_2008} and embedded into minimum bias \PbPb collisions from
the EPOS event generator~\cite{porteboeuf2010producing} tuned with LHC data~\cite{Pierog:2013ria}. Decays of hadronic
particles are described by EvtGen~\cite{Lange:2001uf}, in which final-state radiation is generated using
PHOTOS~\cite{Golonka:2005pn}. The interaction of the generated particles with the detector, and its response,
are implemented using the Geant4 toolkit~\cite{Agostinelli:2002hh,Allison:2006ve} as described in Ref.~\cite{LHCb-PROC-2011-006}.


 \section{Analysis overview}
 \label{sec:analysis}

 The $R_{\Lc/\Dz}$ ratio is obtained through the ratio of corrected yields as:

 	\begin{equation}
   		 R_{\Lc/\Dz}(\pt,y\ {\rm{or}}\ \npart)  = \frac{\BR^{\Dz\rightarrow \Km\pip}}{\BR^{\Lc\rightarrow p\Km\pip}} \frac{Y^{\Lc}(\pt,y\ {\rm{or}}\ \npart)}{Y^{\Dz}(\pt,y\ {\rm{or}}\ \npart)},
    \label{eq:ratio}
 	\end{equation}
 	where $\BR^{\Dz}$ = (3.950 $\pm$ 0.031)\%  ($\BR^{\Lc}$ = (6.28 $\pm$ 0.32)\%) is the branching fraction for the \Dz (\Lc) decay channel~\cite{PDG2020}, respectively; \pt and $y$ are the transverse momentum and rapidity of the \Dz (\Lc) candidate; \npart is the mean number of nucleons participating in the collision and is related to the centrality of the collision; $Y^{\Dz}$ ($Y^{\Lc}$) is the \Dz (\Lc) corrected yield defined as

 	\begin{equation}
   		Y^{\Dz,\Lc}(\pt,y\ {\rm{or}}\ \npart)  = \frac{N^{\Dz,\Lc}(\pt,y\ {\rm{or}}\ \npart)\cdot f^{\Dz,\Lc}_{\rm prompt}(\pt,y\ {\rm{or}}\ \npart)}{\effTot^{\Dz,\Lc}(\pt,y\ {\rm{or}}\ \npart)}.
      \label{eq:yield}
 	\end{equation}

  In Eq.~\ref{eq:yield}, $N^{\Dz,\Lc}$ is the inclusive number of particles measured in the \PbPb dataset, and $f^{\Dz,\Lc}$ is the fraction of particles produced promptly in \PbPb collisions, while $\effTot$ is defined as the total efficiency (see Sec.~\ref{sec:efficiency}). Finally, \npart is defined as the mean number of nucleons participating in the collision. A brief description of the method used to evaluate this quantity is given in Sec.~\ref{sec:cent}.

 	\subsection{Centrality determination}
 	\label{sec:cent}

  In heavy-ion collisions, centrality classes are defined as percentiles of the total inelastic hadronic \PbPb cross-section and are related to the impact parameter $b$ of the collision: the more central (peripheral) the collision, the smaller (greater) the $b$ value. 
  Likewise, \npart increases from peripheral to central collisions.
  The Glauber Monte Carlo (GMC) model~\cite{Loizides:2014vua} is used to estimate all these geometrical quantities from recorded data. A detailed description of the centrality estimation in the LHCb experiment can be found in Ref.~\cite{LHCB-DP-2021-002-001}. The method is based on a binned fit of the total energy deposit in \ecal with the GMC model in MB data, collected with the same trigger conditions as  that of the signal sample. Once the fit is performed, a centrality table is produced, mapping the total \ecal energy deposit and \npart. 

  While the recorded data sample used to fit the GMC model covers the full centrality range, data used to compute $R_{\Lc/\Dz}$ are limited to $N_c$ $<10^4$ and centrality at about $65-90\%$. A one-to-one correspondence between \ecal and the geometrical quantity is performed on an event-by-event basis using the GMC model. Data are divided into three intervals in $N_c$ (1000--3000, 3000--5500 and 5500--10000), based on the statistics available from the signal extraction. For each interval, other quantities (\eg \npart) are derived. A lower cut on $N_c$ and on the total deposited \ecal energy to be above 310 MeV are applied to exclude the centrality range $90-100\%$ where most of the electromagnetic contamination occurs, which could bias the data. Results are given in Table~\ref{tab:centclassvelo}. Three sources of systematic uncertainty associated with \npart of each interval are considered: (i) the reference hadronic cross-section parameter; (ii) the fit uncertainty; (iii) the bin size uncertainty. These uncertainties are summed in quadrature to compute the total systematic uncertainty presented in Table~\ref{tab:centclassvelo}.

  \begin{table}[!tbp]
    \centering
    \caption{values of \npart in $N_c$ intervals, with $\sigma_{\rm part}$ indicating the total systematic uncertainty of \npart.}
    \begin{tabular}{r@{\ --\ }r cc}
    \hline
      \multicolumn{2}{c}{ $N_c$ interval}
      & \npart
      & $\sigma_{\rm part}$\\
      \hline
      1000&10000 &  15.8 & 10.0   \\
      1000&3000 & 6.5 & \ 2.5  \\
      3000&5500 & 12.4 & \ 4.4  \\
      5500&10000 & 26.6 & \ 7.5  \\ 
      \hline
    \end{tabular}
    \label{tab:centclassvelo}
  \end{table}

 	\subsection{Signal extraction}
 	\label{sec:signal}

\begin{figure}[t]
    \centering
    	\includegraphics[width=0.74\textwidth]{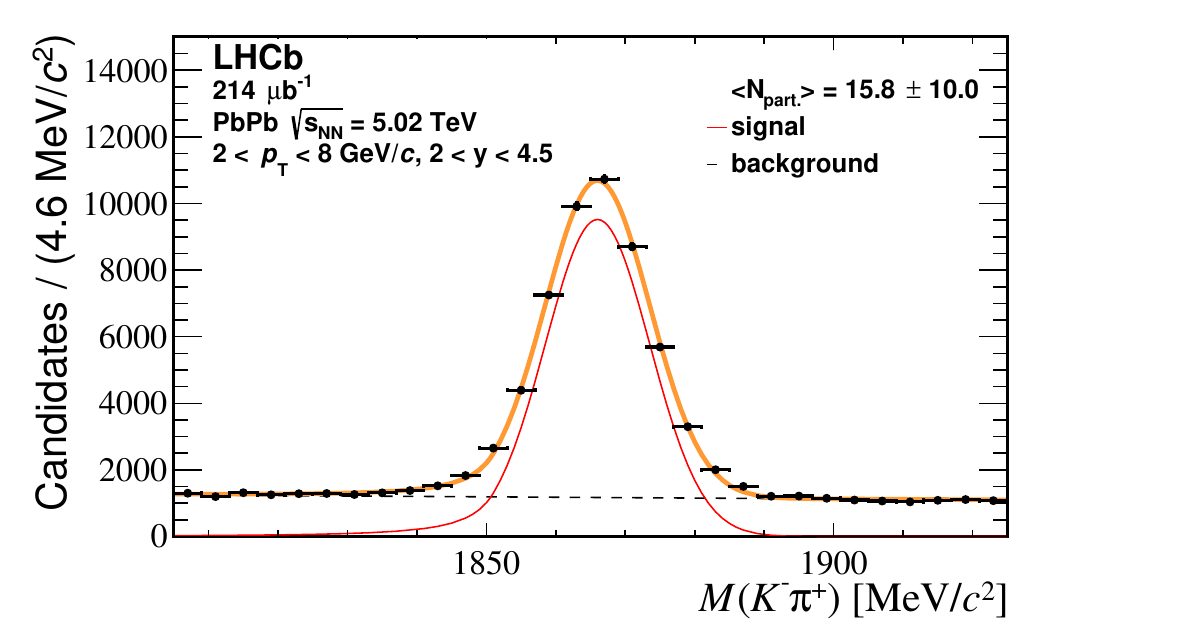}
    	\hfill
    	\includegraphics[width=0.74\textwidth]{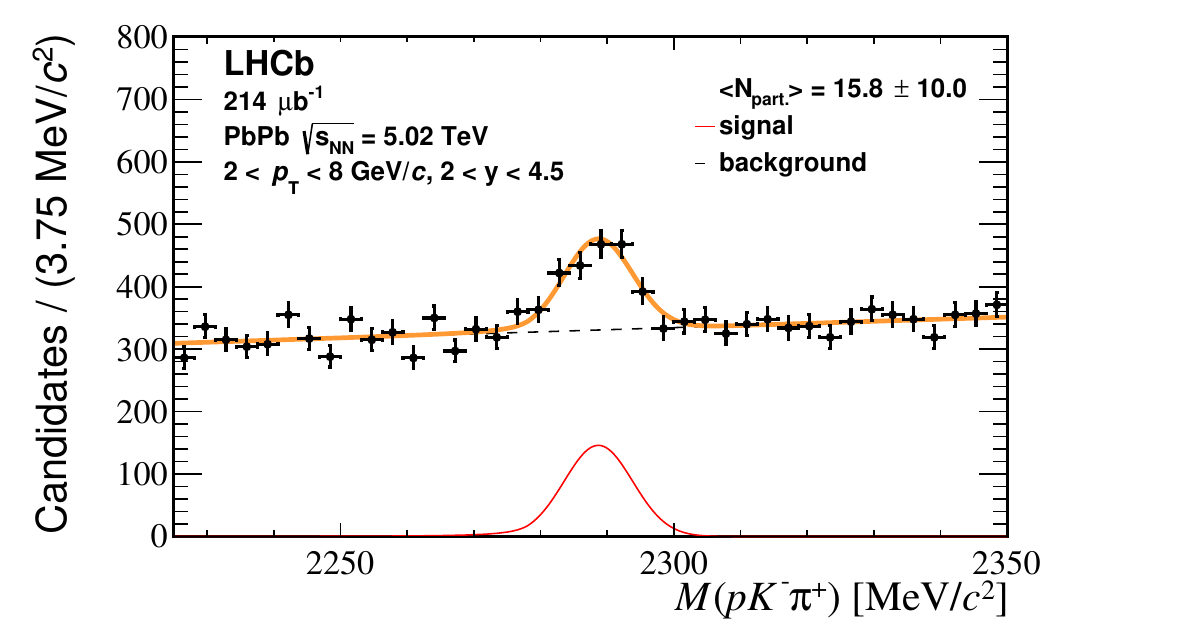}
    	\caption{Invariant-mass spectra of (top) $\Km\pip$ and (bottom) $p\Km\pip$ final states. The data are overlaid with the results of the fit.}
    	\label{fig:minv}
	\end{figure}

  The signal extraction is performed after the selection criteria listed in Sec.~\ref{sec:Detector} are applied. Figure~\ref{fig:minv} shows the (top) $\Km\pip$ and (bottom) $p\Km\pip$ invariant-mass spectra for the selected \Dz and \Lc candidates, respectively. The data are fitted using unbinned maximum-likelihood fits combining a Crystal Ball (CB) function~\cite{Skwarnicki:1986xj} for the signal, and a first-order polynomial function for the background. While the CB function is chosen as it models the radiative tail of the invariant mass peak, the first-order polynomial function is chosen empirically to describe the observed background. The mean and width of the CB function are left free, while the other parameters are fixed to the values extracted from the simulation. An alternative used to assess systematic uncertainty for the background description is to multiply the first-order polynomial by an exponential function. The total number of fitted \Dz (\Lc) signal yield is $46000$ ($600$) events.

\begin{figure}[t]
    \centering
      \includegraphics[width=0.75\textwidth]{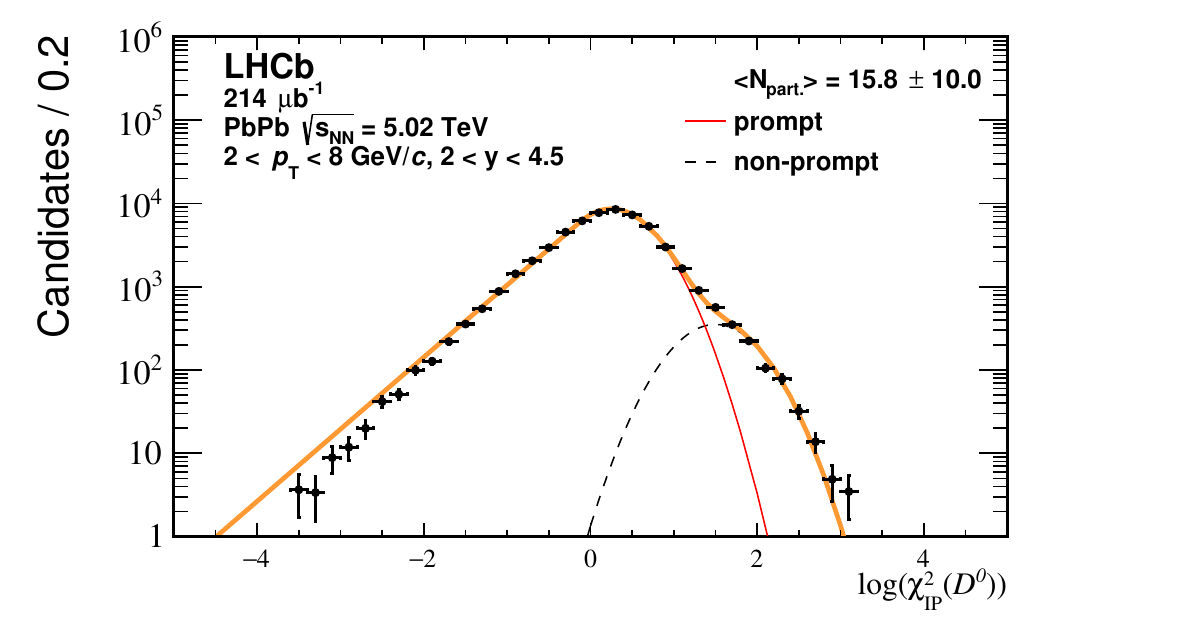}
      \hfill
      \includegraphics[width=0.75\textwidth]{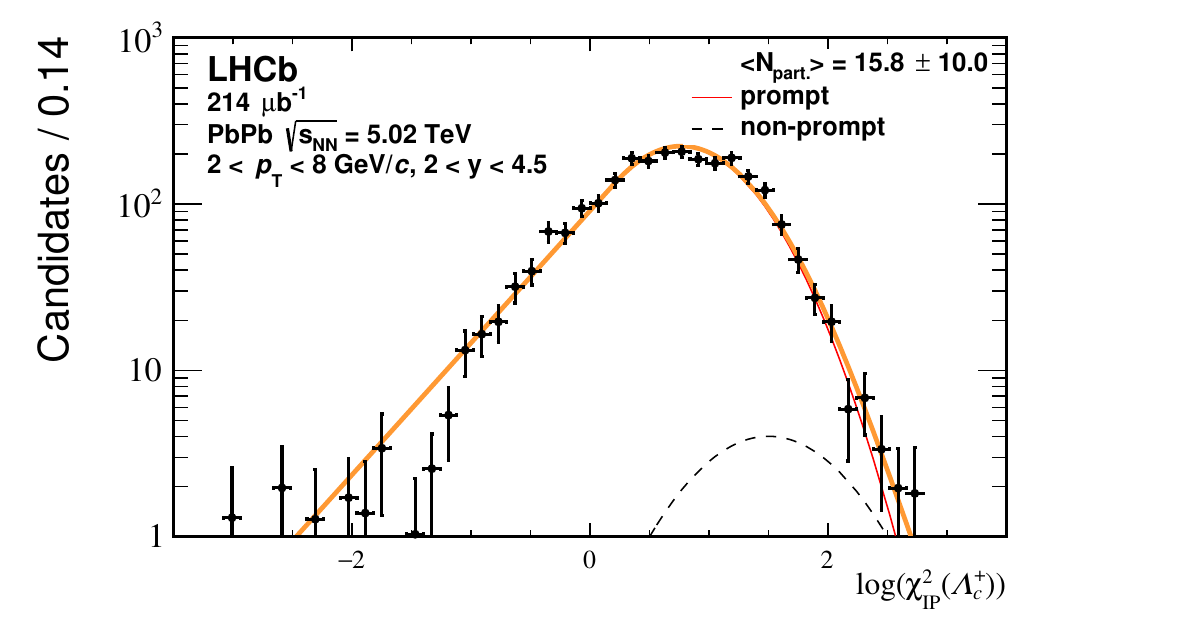}
      \caption{Distributions of  \chisqip\ of the \Dz (top) and \Lc (bottom) after background substraction. The data are overlaid with the results of the fit.}
      \label{fig:fpromt}
  \end{figure}

  To discriminate prompt \Dz (\Lc) from those from a $\bquark$-hadron decay, the background contribution is first subtracted from the datasets using the \sPlot technique~\cite{Pivk:2004ty}. A fit to the $\log(\chisqip)$ distribution of the signal is performed to discriminate the prompt from non-prompt contributions. The \chisqip\ is defined as the difference in the vertex-fit \chisq of a given PV reconstructed with and without the candidate under consideration. An example of such fit is given in Fig.~\ref{fig:fpromt}, where the $\log(\chisqip)$ distributions are fitted with a CB (Gaussian) function for the prompt (non-prompt) component. It is worth noticing that the log(\chisqip) is a better prompt/non-prompt discriminant for \Dz candidates when compared to \Lc candidates. This is due to the shorter lifetime of the \Lc baryon: for non-prompt candidates,  typically the addition of the \Lc track to the PV fit leads to a lower value of \chisqip than that of the non-prompt \Dz track.

 	\subsection{Efficiency estimation}
 	\label{sec:efficiency}

  The total efficiency ($\epsilon_{\rm tot}$) is factorized as 

  \begin{equation}
    \epsilon_{\rm tot}^{\Dz,\Lc}(\pt,y\ {\rm or}\ N_c) = \effAcc^{\Dz,\Lc}(\pt,y)\times\effTrk^{\Dz,\Lc}(\pt,y\ {\rm or\ }N_c)\times\effPID^{\Dz,\Lc}(\pt,y\ {\rm or\ }N_c),
    \label{eq:etot}
  \end{equation}
\noindent
  where $\effAcc$ is the acceptance efficiency; $\effTrk$ is the reconstruction and selection efficiency; $\effPID$ is the PID selection efficiency;  $(N_c,\pt,y)$ indicate given ranges in $N_c$, \pt  and rapidity.

  The acceptance $\effAcc$, for both \Lc and \Dz hadrons, is defined as 

  \begin{equation}
    \effAcc^{\Dz,\Lc}(\pt,y) = \frac{N_{\rm cand}^{2<y<4.5}}{N_{\rm cand}^{\rm fid}},
  \label{eq:acc}
  \end{equation}
\noindent
  where $N_{\rm cand}^{\rm fid}$ are simulated \Dz or \Lc yields within the fiducial acceptance and $N_{\rm cand}^{2<y<4.5}$ indicate the number of the candidates in the fiducial acceptance and having their decay products' rapidity within \mbox{ $2<y<4.5$}. The fiducial acceptance is defined as $2<y<4.5$ and $2<\pt<8$\gevc. This factor is directly computed from simulation based on PYTHIA 8.
  
  The reconstruction and selection efficiency $\effTrk$, for both \Lc and \Dz particles, is defined as:

  \begin{equation}
    \effTrk^{\Dz,\Lc}(\pt,y\ \rm{or}\ N_c) = \frac{N_{\rm cand}^{\rm rec. \& sel.}}{N_{\rm cand}^{2<y<4.5}},
    \label{eq:recnsel}
  \end{equation}
\noindent
  where $N_{\rm cand}^{\rm rec. \& sel.}$ are reconstructed \Dz and \Lc candidates passing the selection criteria within the simulation samples. We recall that the samples are produced with the PYTHIA 8 hard processes embedded into generated MB EPOS events. Several sources of bias are considered for \effTrk due to the limitations of MC to fully model all the bias in the real detector's response. The first source is the tracking algorithm efficiency, defined as the efficiency to reconstruct a track, for which simulations are usually better than in data. Rather than directly measuring the tracking efficiency, the ratio of the efficiency between data and simulation is estimated using two $D^0$ calibration channels ($D^{0}\rightarrow K^{-}\pi^{+}$ and $D^{0} \rightarrow K^{-}\pi^{+}\pi^{-}\pi^{+}$). Their yields are evaluated in \PbPb data and simulation and the difference of their ratio from unity is encoded in a factor $k$. Results are given in Table~\ref{tab:Kfactor}. Other sources of systematic uncertainty are the \textit{ab-initio} assumptions on the \pt, $y$ and $N_c$ distributions, and correlation effects between these variables not accounted for with the embedding technique.
  To account for all these effects, an iterative method based on  data is employed. In the first step, the raw (i.e not corrected for inefficiency) \pt, rapidity, and $N_c$ distributions are extracted from the data based on the signal extraction defined in Sec.~\ref{sec:signal}. In the second step, these distributions are corrected using the data-to-simulation tracking efficiency ($k$ factor) and the PID efficiency (\effPID) computed with the raw kinematic distributions reconstructed in the simulation. In the third step, the reconstructed distributions from the simulation are weighted using several iterations until they match the data as a function of \pt, $y$, and $N_c$ simultaneously. Finally, \effTrk is computed in step four. Steps two to four are repeated until \effTrk converges to a final value, which is the case after three iterations.

  \begin{table}[t!]
    \centering
    \caption{Value of the $k$ factor in intervals of $N_c$. The uncertainties are obtained from the statistical uncertainties of the different \Dz calibration samples in data and MC summed in quadrature.}
    \begin{tabular}{r@{\ --\ }r r@{$\pm$}l}
    \hline
    \multicolumn{2}{c}{ $N_c$ interval}
    & \multicolumn{2}{c}{ $k(N_c)$ } \\
    \hline
    1000&4000  & 0.97 & 0.03  \\
    4000&5500  & 0.93 & 0.04  \\
    5500&10000 & 0.91 & 0.05  \\
    \hline
    \end{tabular}
    \label{tab:Kfactor}
  \end{table}

  The PID efficiency \effPID is computed using the weighted simulation samples. The methodology is similar to that used for \proton\proton collisions~\cite{LHCb-PUB-2016-021}, and is based on a tag-and-probe technique. In this approach, the \effPID for a given probe particle $\effPID^{\rm part}$ (e.g. pion) is computed from a reference sample (e.g. $\Dz \rightarrow \Km\pip$) where a tight selection cut on the tag particle (i.e. the kaon) is applied, while no PID selection is applied to the probe particle. In the next step, the \sPlot technique~\cite{Pivk:2004ty} is used to remove the background with the invariant mass as a discriminating variable. Finally, the PID efficiency of the probe is computed as the fraction of candidates (i.e. \Dz) fulfilling given PID requirement.
  
  Kaon and pion PID efficiencies are computed using the \PbPb\ \Dz sample, while the proton efficiency is computed from $\Lz \rightarrow \pim p$ decays in \PbPb\ data. Two-dimensional maps are then computed for each particle (i.e kaon, pion, proton) as a function of $p$ and $y$, for different ranges in $N_c$. Finally, these maps are used to compute hadron PID efficiency for \Dz and \Lc candidates as
  
    \begin{equation}
    \effPID^{\Dz,\Lc}(\pt,y\ {\rm or}\ N_c) = \frac{\prod^{\rm candidate}_{\rm part}\effPID^{\rm part}(p,y,N_c)}{N_{\rm cand}^{\rm rec. \& sel.}},
    \label{eq:PID}
  \end{equation}
  \noindent
  where $(\pt,y\ {\rm or}\ N_c)$ are the candidates kinematic variables and $\effPID^{\rm part}(p,y,N_c)$ is the single-particle efficiency given as functions of the decay products kinematic variables.
 
 \section{Systematic uncertainties}
 \label{sec:syst}

  Several sources of systematic uncertainty on $R_{\Lc/\Dz}$ are considered. For the signal extraction, three parametrizations of the CB functions are combined with the two background shapes. The systematic uncertainties are taken as the RMS of the results of all the fits for a given bin, considered as uncorrelated between the kinematic intervals. A similar strategy is employed for the prompt fraction estimation, where the Gaussian function is replaced by a Bukin function~\cite{bukin2007fitting}.

 Four sources of systematic uncertainty are considered for the iterative method used to compute \effTrk and \effPID: (i) the uncertainty on the $k$ factor; (ii) the choice of the binning used for the reference raw data distribution; (iii) the sensitivity to the initial reference data distribution; (iv) the uncertainty on the PID maps. The general strategy consists of performing the iterative procedure by varying individually each source of uncertainty within their uncertainties. For each source, \effTrk and \effPID are computed by changing the configuration, and the RMS of all the variations are taken as the systematic uncertainties. For the first source, new results have been obtained on $R_{\Lc/\Dz}$ using 20 values of $k$, varied within uncertainties. The systematic uncertainty associated with the choice of the binning scheme (i.e second source) is evaluated by using a finer scheme than the nominal one. The sensitivity to the initial reference distributions (i.e third source) is tested by evaluating them using the 
 \sPlot technique instead of a fit of invariant-mass spectra. Finally, the uncertainty linked to the PID maps (i.e fourth source) is evaluated using a smearing technique to compute 20 PID maps where the efficiency in each bin is randomly varied within its statistical uncertainty.

 Finally, the last sources of uncertainty considered are the statistical uncertainty coming from the size of the Monte Carlo sample ($\epsilon_{\rm tot}$ (stat.)) after the iterative procedure, and the statistical uncertainty on the acceptance efficiency.

 All the systematic uncertainties are summarized in Table~\ref{tab:uncer-tot}. Each uncertainty category is treated as uncorrelated and is added in quadrature. Systematic uncertainty arising from the \BR ratio of $\Dz\rightarrow \Km\pip$ and $\Lc\rightarrow p\Km\pip$ decays, entering Eq.\ref{eq:ratio}, is also included in Table~\ref{tab:uncer-tot}. This contribution is fully correlated between different kinematic variable intervals.

\begin{table}[ht]
  \centering
  \caption{Summary of the ranges of systematic uncertainties for the considered intervals of $N_c$, \pt and $y$ for the $R_{\Lc/\Dz}$ ratio given in percentage.}
  \begin{adjustbox}{max width=\textwidth}
    \begin{tabular}{l|c|c|c}
      \hline
      Source & $N_c$ & \pt (\gevc) & $y$ \\
      \hline
      $\epsilon_{\rm tot}$ (stat.) & 5.1--7.1 & 4.1--5.1 & 4.1--7.1\\

      Invariant-mass fit & 1--5 & 1--5 & 1--5 \\

      $f_{\rm prompt}$ &  1.8--8.1 & 1.4--10.2 & 3.4--4.6 \\

      $\epsilon_{\rm acc}$ & $<$1& $<$1 & $<$1 \\

      Iterative procedure & 7--9 & 4--11 & 4--8 \\
      \hline
      Total & 9--12 & 8--14 & 8--10\\

      Ratio of decay branching fractions & 5.16 & 5.16 & 5.16\\
      \hline
    \end{tabular}
  \end{adjustbox}
  \label{tab:uncer-tot}
\end{table} 

\clearpage
 \section{Results}
 \label{sec:results}

\begin{figure}[!tbp]
  \centering
  \begin{minipage}{0.49\textwidth}
    \centering
    \includegraphics[width=1.05\textwidth]{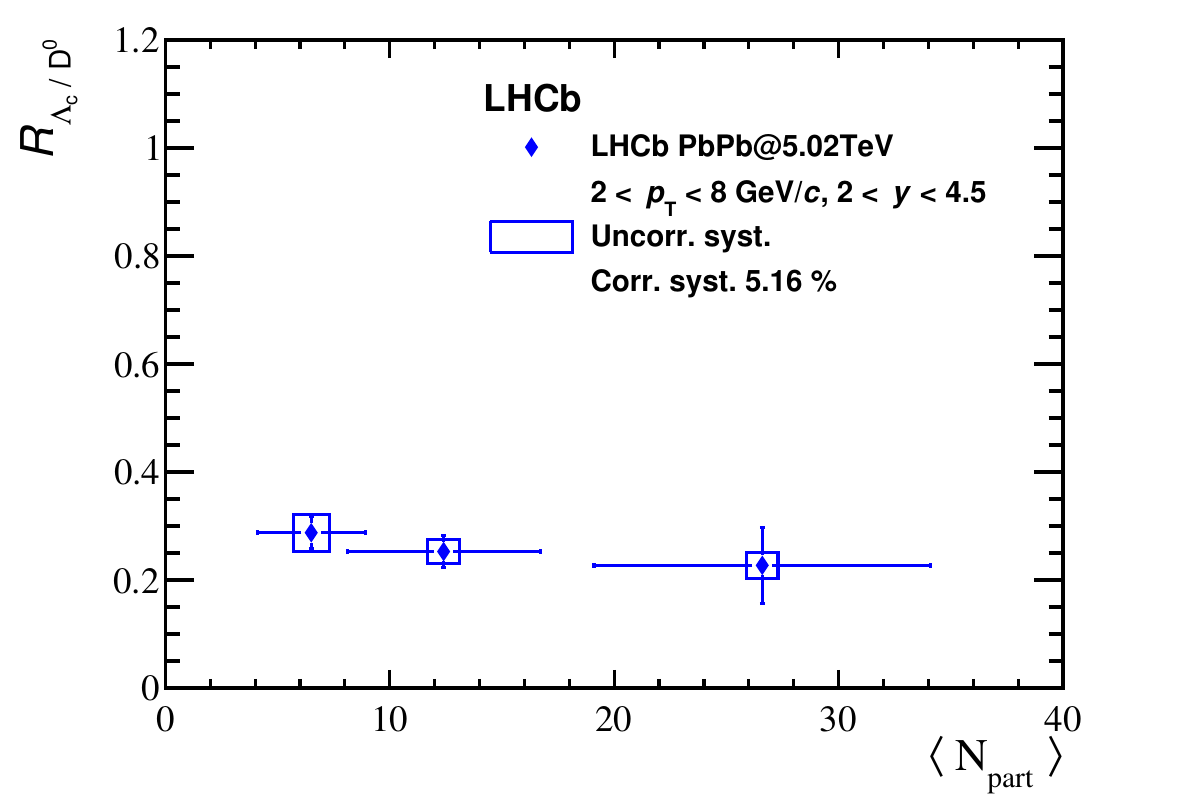}
  \end{minipage}
  \caption{Prompt $R_{\Lc/\Dz}$ production ratios as a function of \npart. The error bars along the y-axis (x-axis) represent the statistical uncertainties (RMS of the \npart value), the boxes around the points the uncorrelated systematic uncertainties.}
  \label{fig:resultnpart}
\end{figure}

\begin{figure}[!tbp]
  \centering
  \begin{minipage}{0.49\textwidth}
    \centering
    \includegraphics[width=1.05\textwidth]{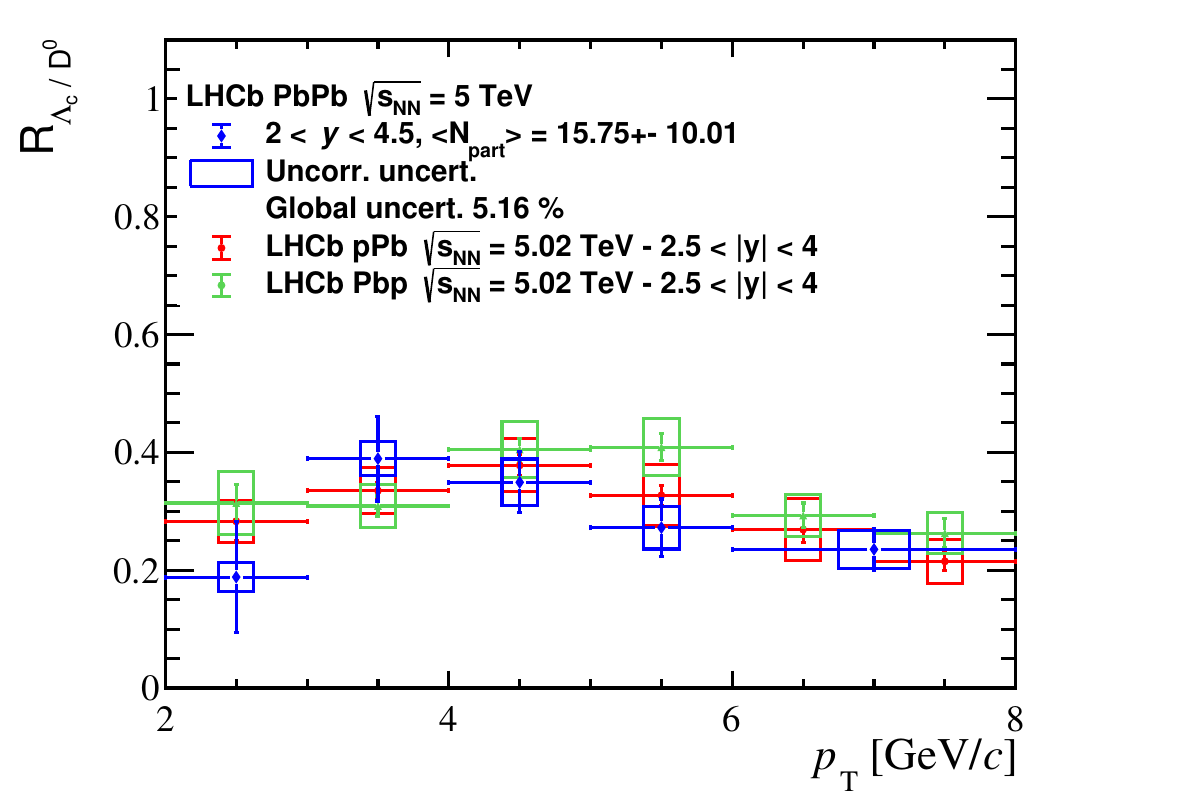}
  \end{minipage}
  \hfill
  \begin{minipage}{0.49\textwidth}
    \centering
    \includegraphics[width=1.05\textwidth]{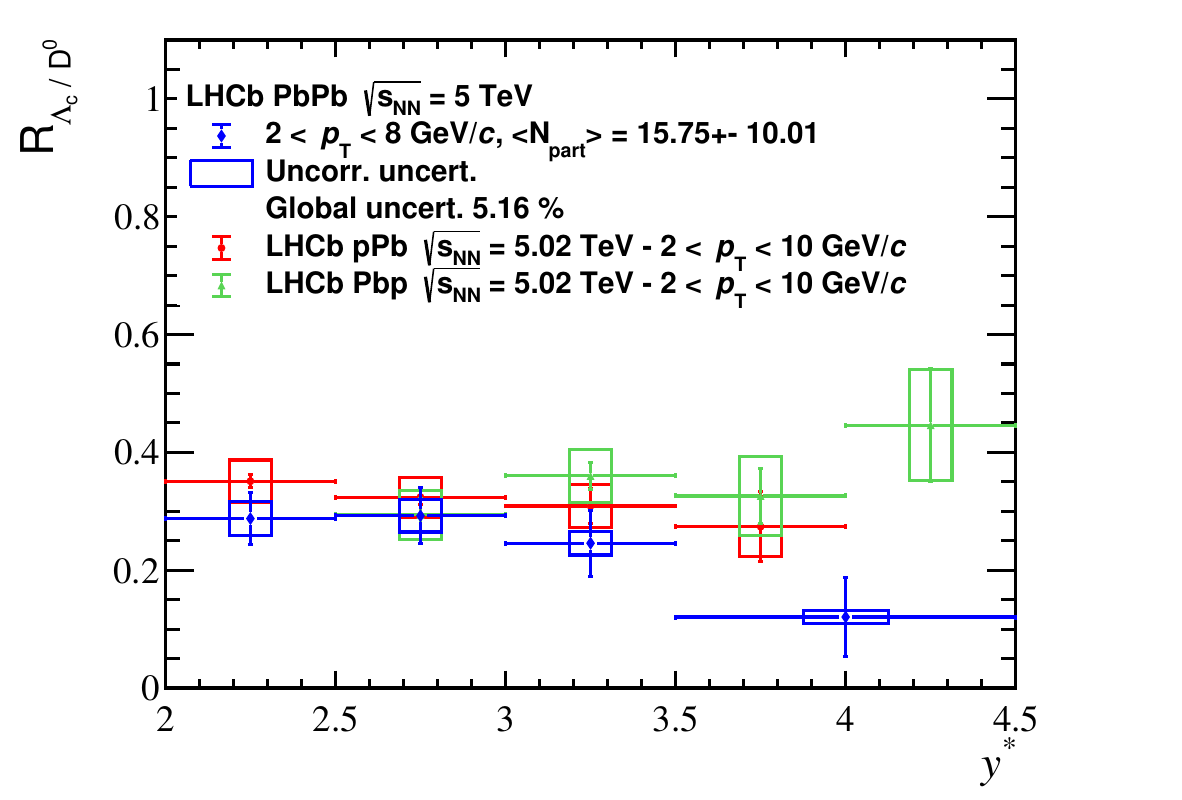}
  \end{minipage}
  \caption{Prompt $R_{\Lc/\Dz}$ production ratios as a function of (left) \pt\ and (right) rapidity, with $y^*$ being the rapidity in the centre-of-mass system for the \pPb and \Pbp results.The error bars represent the statistical uncertainties, the boxes around the points the uncorrelated systematic uncertainties. Results are compared to the same measurement in \pPb and \Pbp collisions at \sqsnn = 5.02 TeV by the LHCb experiment~\cite{LHCb-PAPER-2018-021}.}
  \label{fig:result_ppb}
\end{figure}

\begin{figure}[!tbp]
  \centering
  \begin{minipage}{0.49\textwidth}
    \centering
    \includegraphics[width=1\textwidth]{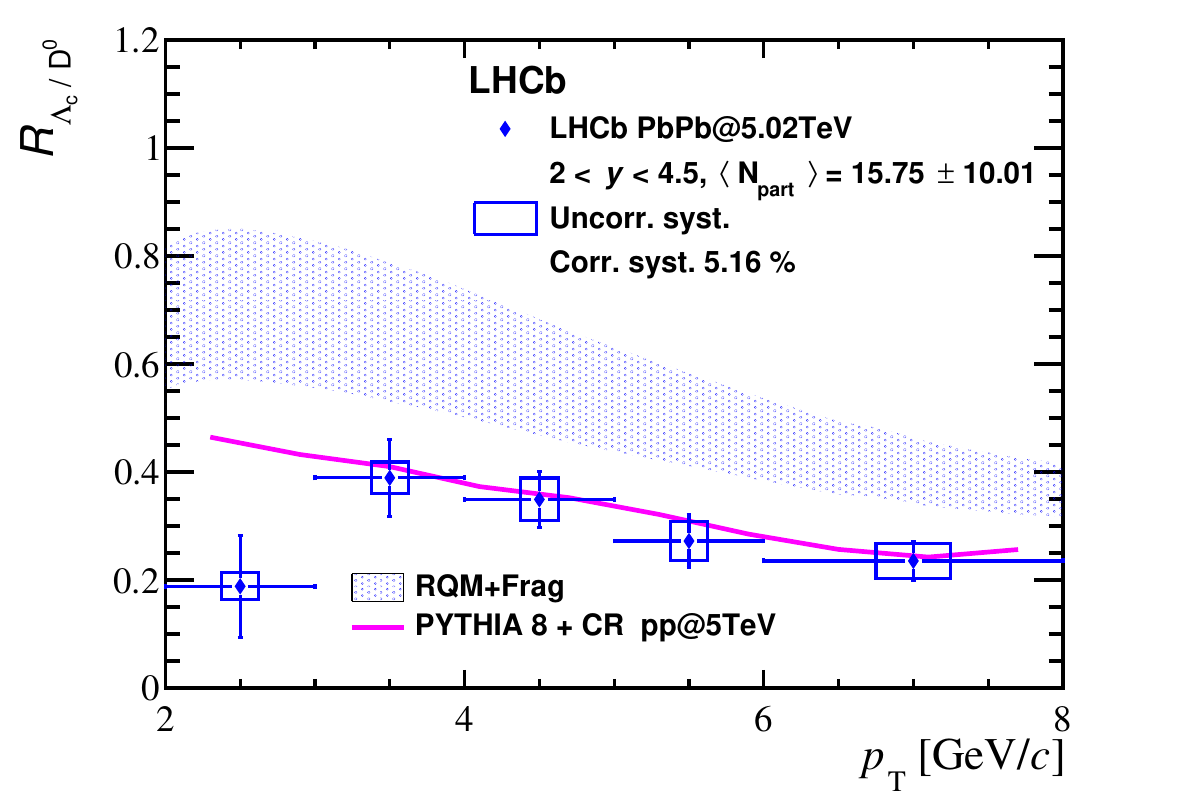}
  \end{minipage}
  \hfill
  \begin{minipage}{0.49\textwidth}
    \centering
    \includegraphics[width=1\textwidth]{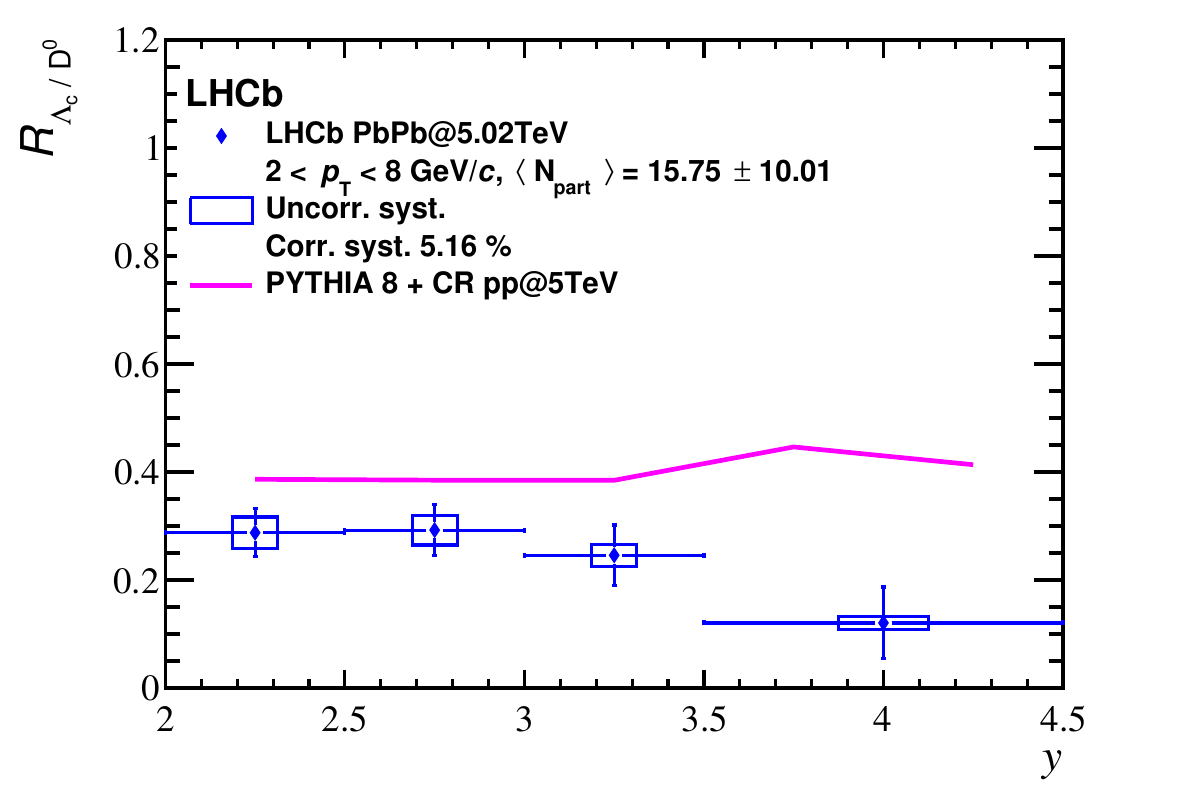}
  \end{minipage}
  \caption{Prompt $R_{\Lc/\Dz}$ production ratios as a function of (left) \pt\ and (right) rapidity. The error bars represent the statistical uncertainties, the boxes around the points the uncorrelated systematic uncertainties. Results are compared to PYTHIA 8~\cite{Christiansen2015} in \proton\proton collisions at \sqs = 13 TeV (magenta band) and predictions from a Statical Hadronization Model~\cite{He_2019} (blue band).}
  \label{fig:result_th}
\end{figure}

Results for the $R_{\Lc/\Dz}$ production ratio are given in Table~\ref{tab:resultsfinal}.
The $N_c$ variable is replaced by \npart as shown in Table~\ref{tab:centclassvelo}.
As can be seen in Fig.~\ref{fig:resultnpart}, the \pt-integrated $R_{\Lc/\Dz}$ shows no dependence on \npart, within uncertainties, with a mean value $\langle R_{\Lc/\Dz} \rangle \sim 0.27$.
In Fig.~\ref{fig:result_ppb}, the \pt\ and $|y^*|$ dependent $R_{\Lc/\Dz}$ results are compared to those from \pPb data at \mbox{ $\sqsnn = 5.02 \tev$}~\cite{Aaij:2018iyy}, showing a good agreement between the measurements. Here $|y^*|$ is defined as the absolute value of rapidity in the center-of-mass system ($y^*$). Due to the beam boost, the \pPb (proton beam towards LHCb) and \Pbp (lead beam towards LHCb) data were recorded at forward ($1.5 < y^* < 4.0$) and backward ($-4.5 < y^* < -2.5$) rapidity in the laboratory rest frame, respectively.
This observation is consistent with the fact that the two samples have relatively close center-of-mass energy, and similar \npart values ($\npart \sim 7.9$ in \pPb collisions at \sqsnn = 5.02 \tev as measured by ALICE~\cite{ALICE:2014xsp}).
The same measurements of $R_{\Lc/\Dz}$ versus \pt and $y$ are compared to theoretical predictions in Fig.~\ref{fig:result_th}.
Both dependencies are compared to predictions from PYTHIA 8~\cite{Christiansen2015} in \proton\proton collisions at \mbox{\sqs = 5.02 \tev} using the beyond-the-leading-color contributions.
For these predictions, a special tuning is used to increase the \Lc color-reconnection mechanism at the expense of \D mesons (see Appendix~\ref{sec:pythiatuning}). In addition, the CR mechanism is also allowed.
A good description of the \pt trend is found between theory and data for \mbox{ $\pt > 3 \gevc$}, while tensions are observed at $\pt< 3 \gevc$. The constant trend versus rapidity is reproduced by theory predictions within three standard deviations although systematically higher.
The \pt dependence is also compared to predictions from the SHM~\cite{He_2019} for which an augmented set of excited charm-baryon states decaying into \Lc is considered, and where fragmentation functions are used to compute the charm-hadron \pt spectra. The uncertainty band encompasses the uncertainty from varying the branching fractions from 50\% to 100\%. According to this model, these additional states could explain the large $R_{\Lc/\Dz}$ observed by the ALICE experiment at mid-rapidity in \proton\proton and \pPb collisions at \sqsnn = 5.02 \tev~\cite{Acharya_2019,ALICE:2020wla}. These predictions are systematically higher than the LHCb data versus \pt.



  \begin{table}[b]
    \centering
    \caption{Results for the prompt $R_{\Lc/\Dz}$ production ratio, where the first uncertainties are statistical and the second systematic. A fully correlated systematic uncertainty of 5.16\%, due to the limited knowledge of the external decay branching fractions, affects all the intervals.}
    \begin{adjustbox}{max width=\textwidth}
    \begin{tabular}{c|c}
      \hline
    \pt(GeV/$c$) & $R_{\Lc/\Dz}$\\
      \hline
        ~2\ --\ 3 & 0.188 $\pm$ 0.095 $\pm$ 0.025 \\
        
        ~3\ --\ 4 & 0.389 $\pm$ 0.072  $\pm$ 0.029 \\
        
        ~4\ --\ 5 & 0.349 $\pm$ 0.052  $\pm$ 0.040 \\
        
        ~5\ --\ 6 & 0.272 $\pm$ 0.049 $\pm$ 0.036 \\
        
        ~6\ --\ 8 & 0.235 $\pm$ 0.035  $\pm$ 0.032 \\
        \hline
    $y$ & $R_{\Lc/\Dz}$ \\
        \hline
        ~2.0\ --\ 2.5 & 0.288 $\pm$ 0.044  $\pm$ 0.029 \\
        
        ~2.5\ --\ 3.0 & 0.292 $\pm$ 0.048  $\pm$ 0.028 \\
        
        ~3.0\ --\ 3.5 & 0.246 $\pm$ 0.056  $\pm$ 0.020 \\
        
        ~3.5\ --\ 4.5 & 0.120 $\pm$ 0.067  $\pm$ 0.011 \\
        \hline
      \npart\ $\pm$ $\sigma_{\rm part}$~~ & $R_{\Lc/\Dz}$ \\
        \hline
        \ 6.5 $\pm\ 2.5$ & 0.288 $\pm$ 0.029 $\pm$ 0.034 \\
        
        12.4 $\pm\ 4.4$& 0.253 $\pm$ 0.029 $\pm$ 0.022 \\
        
        26.6 $\pm\ 7.5$& 0.227 $\pm$ 0.071  $\pm$ 0.024 \\
        \hline
    \end{tabular}
    \end{adjustbox}
    \label{tab:resultsfinal}
  \end{table}

  \clearpage

\clearpage
 \section{Conclusions}
 This paper reports the first measurements of the $R_{\Lc/\Dz}$ production cross-section ratio in peripheral \PbPb collisions at \sqsnn = 5.02 \tev by the LHCb experiment. The $R_{\Lc/\Dz}$ shows no significant dependence on either rapidity or \npart in peripheral collisions within uncertainties and has a mean value of $\langle R_{\Lc/\Dz} \rangle \sim 0.27$. However, the ratio tends to decrease at lower \pt. More data are needed to confirm the results. Results are consistent with previous LHCb measurements in \pPb collisions at \sqsnn = 5.02 \tev~\cite{LHCb-PAPER-2018-021}. Compared to theory predictions, the results are compatible within one standard deviation with the PYTHIA 8 predictions in \proton\proton collisions at \sqs = 5.02 \tev, including the CR mechanism at next-to-leading color approximation, except at low \pt. In contrast, a systematic discrepancy versus \pt is observed with the SHM model predictions with an extended set of baryon states. These new experimental results point toward a strong dependence of $R_{\Lc/\Dz}$ on rapidity when compared to ALICE measurements at mid-rapidity in \proton\proton and \pPb collisions~\cite{ALICE:2020wla}, which could help to constrain theory predictions in this particular phase-space region.
 \label{sec:conclusion}

 \clearpage


\clearpage
\newpage

\section*{Additional material}

\appendix

\section{PYTHIA 8 Tuning}
\label{sec:pythiatuning}

The Pythia tuning represents the code that produced the results presented in Fig.~\ref{fig:result_th}.

\begin{table}[h]
    \centering
    \begin{tabular}{l|l}
            \hline
    Parameter & Value \\
        \hline
        \hline
    SoftQCD:all &on \\
    StringFlav:probQQ1toQQ0join &0.0275,0.0275,0.0275,0.0275 \\
    StringFlav:probQQtoQ &0.0780000 \\
    StringFlav:ProbStoUD &0.2 \\
    StringZ:aLund &0.36 \\
    StringZ:bLund &0.56 \\
    StringFlav:mesonCvector &1.35 \\
    ColourReconnection:mode &1 \\
    BeamRemnants:remnantMode &1 \\
    ColourReconnection:allowDoubleJunRem &off \\
    MultipartonInteractions:pT0Ref &2.15 \\
    PartonVertex:ProtonRadius &0.7 \\
    PartonVertex:setVertex &on \\
    Ropewalk:beta &0.1 \\
    Ropewalk:deltat &0.05 \\
    Ropewalk:doFlavour &on \\
    Ropewalk:doShoving &on \\
    Ropewalk:gAmplitude &0.0 \\
    Ropewalk:RopeHadronization &on \\
    Ropewalk:tShove &0.1 \\
            \hline
    \end{tabular}
    \caption{PYTHIA 8 tuning used for the theory predictions.}
    \label{tab:pythia_param}
\end{table}

\newpage

\section{Comparison to other experiments}

In this appendix, the results from this paper are compared to measurements from other experiments.

\begin{figure}[h]
  \centering
  \begin{minipage}{0.7\textwidth}
    \centering
    \includegraphics[width=1\textwidth]{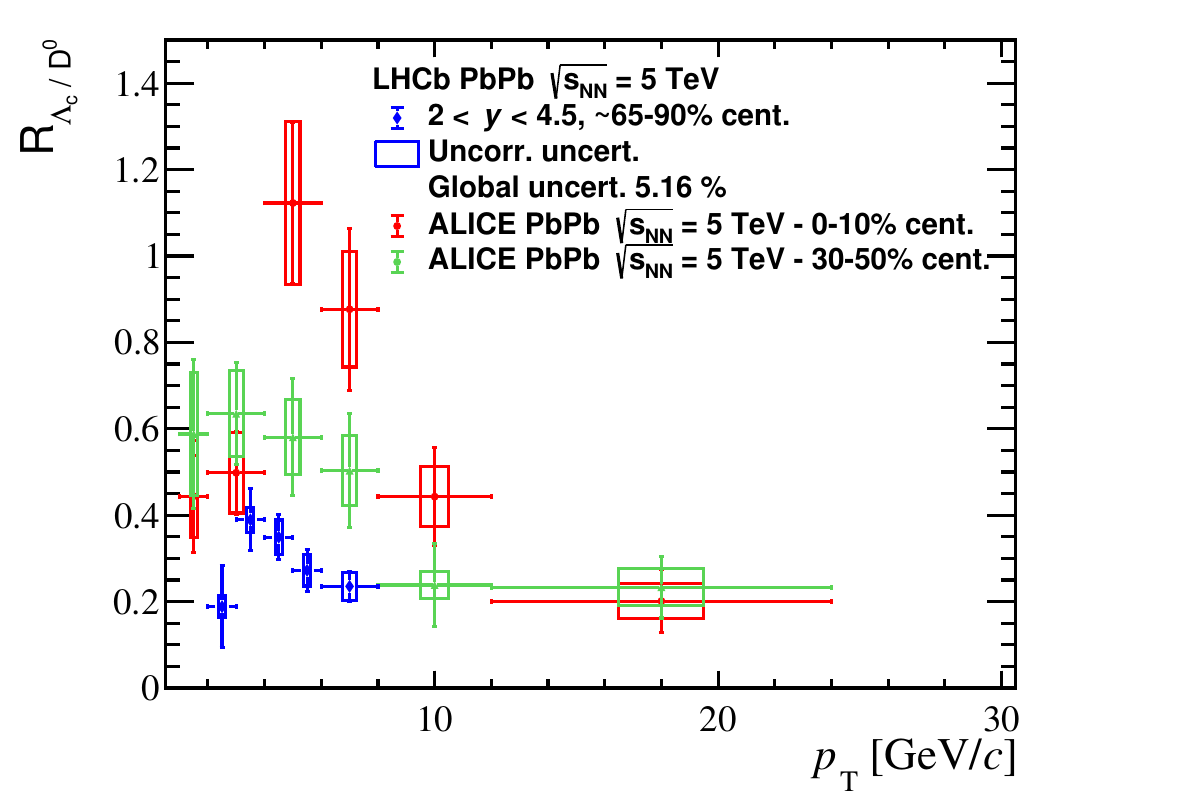}
  \end{minipage}
  \hfill
  \begin{minipage}{0.7\textwidth}
    \centering
    \includegraphics[width=1\textwidth]{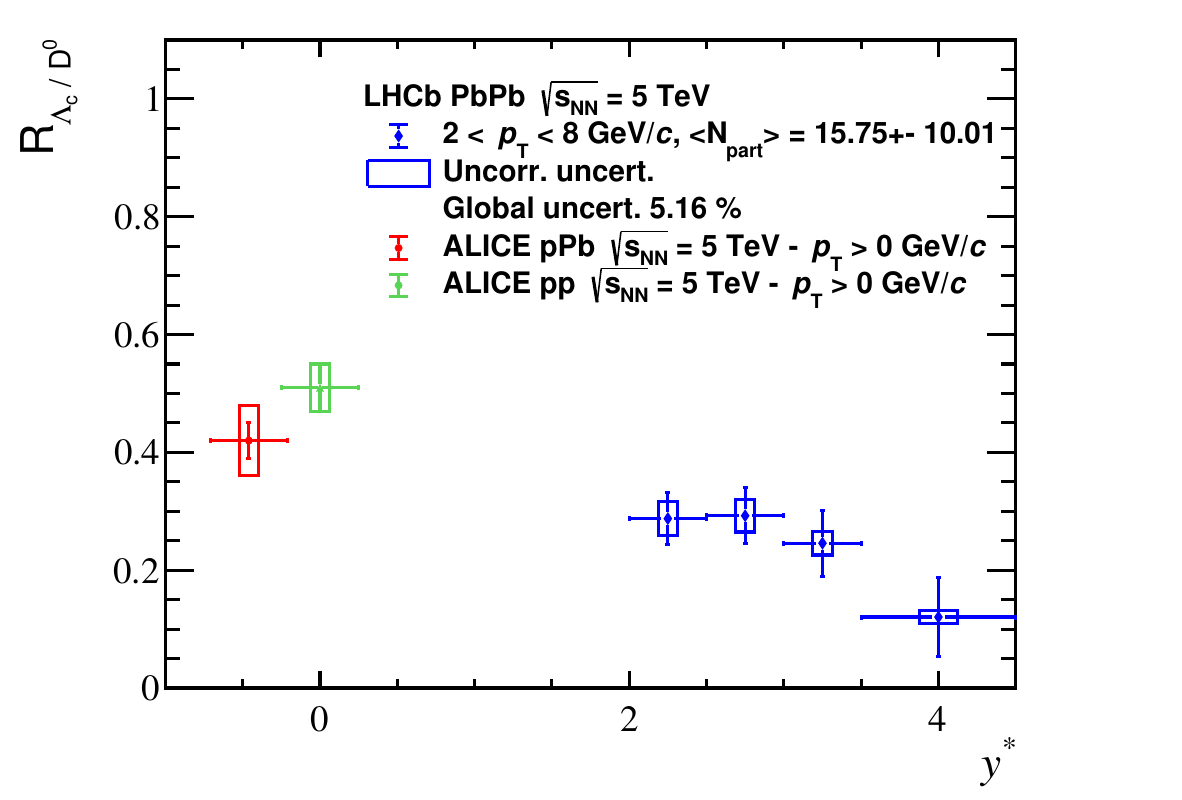}
  \end{minipage}
  \caption{Prompt $R_{\Lc/\Dz}$ production ratios as a function of \pt (top panel) and rapidity in the centre-of-mass system (bottom panel). The error bars represent the statistical uncertainties, the boxes around the points the uncorrelated systematic uncertainties. Results are compared to the ALICE measurements (brown) in \proton\proton and (light blue) \pPb collisions at \sqsnn = 5.02 TeV~\cite{ALICE:2020wla}.}
  \label{fig:result_alice}
\end{figure}

\begin{figure}[h]
  \centering
  \begin{minipage}{0.7\textwidth}
    \centering
    \includegraphics[width=1\textwidth]{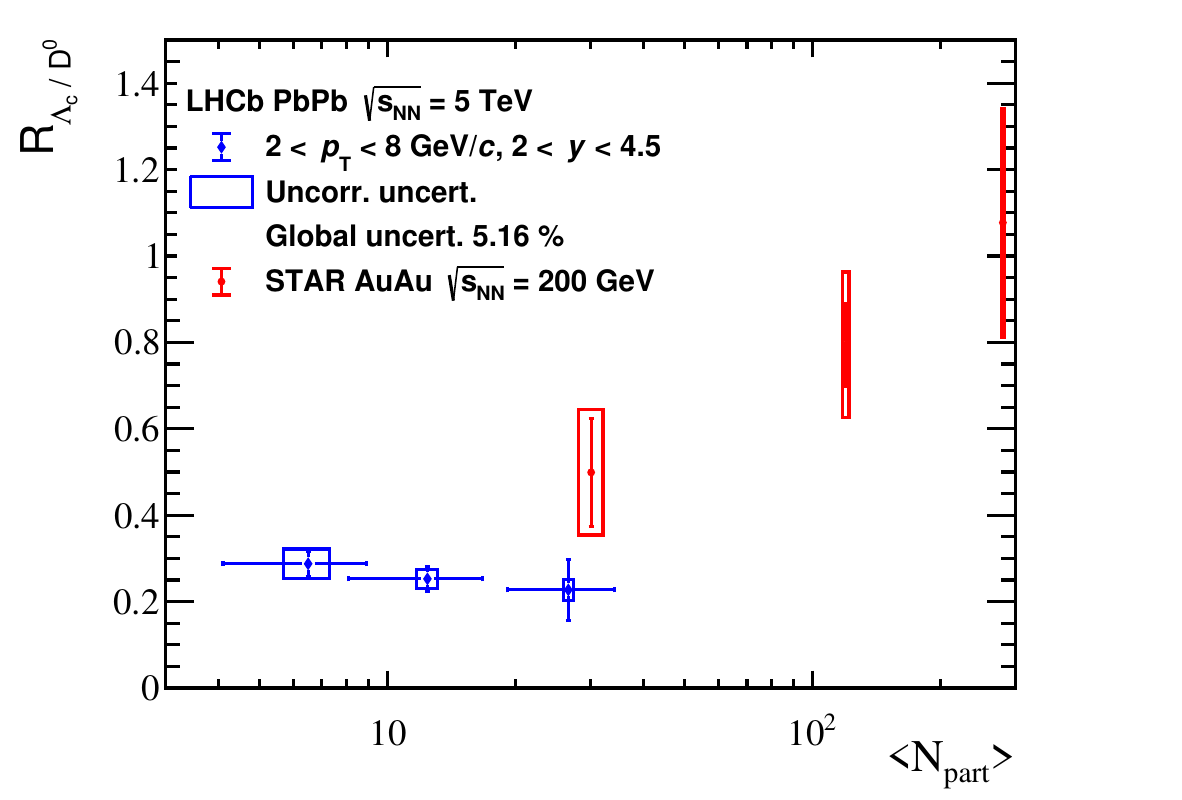}
  \end{minipage}
  \caption{Prompt $R_{\Lc/\Dz}$ production ratios as a function of \npart. The error bars represent the statistical uncertainties, the boxes around the points the uncorrelated systematic uncertainties. Results are compared to the STAR measurements in Au-Au collisions at \sqsnn = 200 GeV~\cite{Adam:2019hpq}.}
  \label{fig:result_star}
\end{figure}
\clearpage
\newpage

\addcontentsline{toc}{section}{References}
\bibliographystyle{LHCb}
\bibliography{main,standard,LHCb-PAPER,LHCb-CONF,LHCb-DP,LHCb-TDR}

\newpage
\centerline
{\large\bf LHCb collaboration}
\begin
{flushleft}
\small
R.~Aaij$^{32}$\lhcborcid{0000-0003-0533-1952},
A.S.W.~Abdelmotteleb$^{50}$\lhcborcid{0000-0001-7905-0542},
C.~Abellan~Beteta$^{44}$\lhcborcid{0009-0009-0869-6798},
F.~Abudin{\'e}n$^{50}$\lhcborcid{0000-0002-6737-3528},
T.~Ackernley$^{54}$\lhcborcid{0000-0002-5951-3498},
B.~Adeva$^{40}$\lhcborcid{0000-0001-9756-3712},
M.~Adinolfi$^{48}$\lhcborcid{0000-0002-1326-1264},
H.~Afsharnia$^{9}$,
C.~Agapopoulou$^{13}$\lhcborcid{0000-0002-2368-0147},
C.A.~Aidala$^{76}$\lhcborcid{0000-0001-9540-4988}
% ...

{\footnotesize \it

% ...
$^{9}$Universit{\'e} Clermont Auvergne, CNRS/IN2P3, LPC, Clermont-Ferrand, France\\
% ...
$^{13}$LPNHE, Sorbonne Universit{\'e}, Paris Diderot Sorbonne Paris Cit{\'e}, CNRS/IN2P3, Paris, France\\
% ...
$^{32}$Nikhef National Institute for Subatomic Physics, Amsterdam, Netherlands\\
% ...
$^{40}$Instituto Galego de F{\'\i}sica de Altas Enerx{\'\i}as (IGFAE), Universidade de Santiago de Compostela, Santiago de Compostela, Spain\\
% ...
$^{44}$Physik-Institut, Universit{\"a}t Z{\"u}rich, Z{\"u}rich, Switzerland\\
% ...
$^{48}$H.H. Wills Physics Laboratory, University of Bristol, Bristol, United Kingdom\\
% ...
$^{50}$Department of Physics, University of Warwick, Coventry, United Kingdom\\
% ...
$^{54}$Oliver Lodge Laboratory, University of Liverpool, Liverpool, United Kingdom\\
% ...
$^{62}$Syracuse University, Syracuse, NY, United States\\
% ...
$^{76}$University of Michigan, Ann Arbor, MI, United States, associated to $^{62}$\\
\bigskip
% ...
}
\end{flushleft}

\end{document}